%% file: paper.tex
\documentclass[conference,10pt]{IEEEtran}
\IEEEoverridecommandlockouts
\usepackage{cite}
\usepackage{amsmath,amssymb,amsfonts}
\usepackage{algorithmic}
\usepackage{graphicx}
\usepackage{textcomp,balance}
\def\BibTeX{{\rm B\kern-.05em{\sc i\kern-.025em b}\kern-.08em
    T\kern-.1667em\lower.7ex\hbox{E}\kern-.125emX}}
\usepackage{multicol}

\usepackage{dblfloatfix}

\newif\ifnewr
\newif\ifoldr

\newrtrue
\usepackage{amsmath,graphicx,url,times,amssymb,url}
\usepackage{color}
\usepackage{siunitx}
\usepackage[table]{xcolor}
\usepackage{collcell}
\usepackage{hhline}
\usepackage{pgf}
\usepackage{multirow}
\usepackage{booktabs, acronym, soul, enumitem}

\usepackage{subcaption, caption}

\def\colorModel{hsb} %You can use rgb or hsb

\newcommand\ColCell[1]{
  \pgfmathparse{#1<100?1:0}  %Threshold for changing the font color into the cells
    \ifnum\pgfmathresult=0\relax\color{white}\fi
  \pgfmathsetmacro\compA{0}      %Component R or H
  \pgfmathsetmacro\compB{#1/100} %Component G or S
  \pgfmathsetmacro\compC{1}      %Component B or B
  \edef\x{\noexpand\centering\noexpand\cellcolor[\colorModel]{\compA,\compB,\compC}}\x #1
  } 
\newcolumntype{E}{>{\collectcell\ColCell}m{0.5cm}<{\endcollectcell}}  %Cell width
\newcommand*\rot{\rotatebox{90}}

\newcommand\CorCell[1]{
  \pgfmathparse{#1<1000?1:0}  %Threshold for changing the font color into the cells
    \ifnum\pgfmathresult=0\relax\color{white}\fi
  \pgfmathsetmacro\compA{0}      %Component R or H
  \pgfmathsetmacro\compB{#1/265} %Component G or S
  \pgfmathsetmacro\compC{1}      %Component B or B
  \edef\x{\noexpand\centering\noexpand\cellcolor[\colorModel]{\compA,\compB,\compC}}\x #1
  } 
\newcolumntype{F}{>{\collectcell\CorCell}m{0.5cm}<{\endcollectcell}}  %Cell width

\newacro{STFT}[STFT]{Short Time Fourier Transform}
\newacro{MSC}[MSC]{Magnitude Squared Coherence}
\newacro{GCC}[GCC]{Generalized Cross-Correlation}
\newacro{DFT}[DFT]{Discrete Fourier Transform}
\newacro{DOA}[DoA]{Direction of Arrival}
\newacro{PSF}[PSF]{Phase Sensitive Filter}
\newacro{MAE}[MAE]{Mean Absolute Error}
\newacro{MSE}[MSE]{Mean Squared Error}
\newacro{LSTM}[LSTM]{Long Short-Term Memory}
\newacro{SIR}[SIR]{Signal to Interference Ratio}
\newacro{SNR}[SNR]{Signal to Noise Ratio}
\newacro{SSNR}[SSNR]{Segmental Signal to Noise Ratio}
\newacro{STOI}[STOI]{Short-Term Objective Intelligibility}
\newacro{RIR}[RIR]{Room Impulse Response}
\newacro{DSB}[DSB]{Delay and Sum Beamforming}
\newacro{SRP}[SRP]{Steered Response Power}
\newacro{SRP-PHAT}[SRP-PHAT]{\ac{SRP} with phase transform}
\newacro{GCC-PHAT}[GCC-PHAT]{\ac{GCC} with phase transform}

\newacro{ASR}[ASR]{Automatic Speech Recognition}
\newacro{ASL}[ASL]{Acoustic Source Localization}
\newacro{SSL}[SSL]{Sound Source Localization}
\newacro{MVDR}[MVDR]{Minimum Variance Distortionless Response}
\newacro{WGN}[WGN]{White Gaussian Noise}
\newacro{RTF}[RTF]{Relative Transfer Function}
\newacro{TDOA}[TDoA]{Time Difference of Arrival}
\newacro{VAD}[VAD]{Voice Activity Detection}
\newacro{TF}[TF]{Time-Frequency}
\newacro{DRR}[DRR]{Direct to Reverberant Ratio}
\newacro{RNN}[RNN]{Recurrent Neural Network}
\newacro{DNN}[DNN]{Deep Neural Network}
\newacro{CNN}[CNN]{Convolutional Neural Network}
\newacro{CRNN}{Convolutional Recurrent Neural Network}
\newacro{ReLU}[ReLU]{Rectified Linear Unit}

\newcommand{\ssection}[1]{\vspace{-0.0mm}\subsubsection{#1}\vspace{-0.0mm} }

\newcommand{\ssubsection}[1]{\vspace{0mm}\subsection{#1}\vspace{0mm} }

\makeatletter
\patchcmd{\@maketitle}
  {\addvspace{0.5\baselineskip}\egroup}
  {\addvspace{-1\baselineskip}\egroup}
  {}
  {}
\makeatother

\usepackage{booktabs, graphicx}
\begin{document}
% Title.
% --------------------
%\title{Speech Detection and Localization in Diverse Everyday Environments with a Mobile Microphone Array}
\title{Mobile Microphone Array Speech Detection and Localization in Diverse Everyday Environments\vspace{-2mm}}

%%%%% Centre for Immersive Visual Technologies
%\vspace{-21mm}
\author{
\IEEEauthorblockN{1\textsuperscript{st} Pasi Pertil\"a }
\IEEEauthorblockA{Tampere University, Finland\\\textnormal{ pasi.pertila@tuni.fi}}
%\hspace{4cm}
\and
\IEEEauthorblockN{2\textsuperscript{nd} Emre Cak{\i}r$^{\dagger}$\thanks{\vspace{-3mm} ${}^\dagger$Affiliation during research. Current affiliation Inscripta Oy, Finland.}}
\IEEEauthorblockA{Tampere University, Finland \\
\textnormal{emre.cakir@inscripta.io}}
%\hspace{2cm}
\and
\IEEEauthorblockN{3\textsuperscript{rd} Aapo Hakala }
\IEEEauthorblockA{Tampere University,  Finland \\ \textnormal{aapo.hakala@tuni.fi }}
%\hspace{2cm}
\and
\IEEEauthorblockN{4\textsuperscript{th} Eemi Fagerlund }
\IEEEauthorblockA{Tampere University,   Finland\\ \textnormal{eemi.fagerlund@tuni.fi  }}
%\hspace{2cm}
\and
\IEEEauthorblockN{5\textsuperscript{th} Tuomas Virtanen }
\IEEEauthorblockA{Tampere University,  Finland\\ \textnormal{tuomas.virtanen@tuni.fi }}
%\hspace{2cm}
\and
\IEEEauthorblockN{7\textsuperscript{th} Archontis Politis }
\IEEEauthorblockA{Tampere University,  Finland\\ \textnormal{archontis.politis@tuni.fi }}
%\hspace{2cm}
\and
\IEEEauthorblockN{8\textsuperscript{th} Antti Eronen }
\IEEEauthorblockA{Nokia Technologies Oy\\
\textnormal{antti.eronen@nokia.com}}
}
\maketitle

%%%%

%% Single addresses (uncomment and modify for single-address case).
%% --------------------
%\name{Author(s) Name(s)\thanks{Thanks to XYZ agency for funding.}}
%\address{Author Affiliation(s)}
%%
%% For example:
%% ------------
%%\address{School\\
%%       Department\\
%%       Address}

% Two addresses
% --------------------
%\twoauthors
%  {John Doe\sthanks{Thanks to ABC agency for funding.}}
%    {Fictional University\\
%Computer Science Dept., 2133 Long Road\\
%     Gotham, NY 10027, USA \\
%     john@fictional.edu}
%  {Maria Ortega\sthanks{Thanks to XYZ agency for funding.}}
%    {University of the Imagination \\
%     Big Engineering Building, 8765 Dream Blvd. \\
%     New Chicago, IL 60626, USA \\
%     maria@imagination.edu}

% Many authors with many addresses
% --------------------
%\name{Pasi Pertil\"a$^{1,2}, $%\sthanks{Thanks to ABC agency for funding.}
%      Emre Cakir$^{1}$,%\sthanks{Thanks to XYZ agency for funding.}
%      Aapo Hakala$^{1}$,% \sthanks{Also many thanks.}
%      Eemi Fagerlund$^{1}$,  
%      Tuomas Virtanen$^{1}$, and
%      Antti Eronen$^{3}$
%      }
%\address{Tampere University, Faculty of Communication Sciences and Information Technology, \\
%$^1$Audio Research Group, 
%$^2$Centre for Immersive Visual Technologies  \\ Tampere, Finland  \\
%$^3$Nokia Technologies Oy, Tampere, Finland
%}

%\ninept
%\maketitle

%\begin{sloppy}

\begin{abstract}
\input abstract.tex
\end{abstract}

%\begin{keywords}
%Sound Event Detection, Sound Source Localization, Voice Activity Detection
%\end{keywords}

\section{Introduction}
\label{sec:intro}
\input introduction.tex

\vspace{-1mm}
\section{Dataset}
\label{sec:data}
\vspace{-1mm}

\input dataset.tex

\vspace{-0mm}
\section{Hierarchical Classification Approach}
\vspace{-0mm}
\label{sec:method}
%
%The used labels for sound event detection and localization are here defined as: "speech front", "speech back", "something else", and "background".
Given the described labeling scheme, the direct approach is a flat classifier with three multilabel binary output values. The outputs are the presence probabilities for labels "speech back", "speech front" and "something else". To solve the task, the model's input features should contain both spectral and directional information. However, sound event characteristics and direction properties are two very different types of information and therefore challenging to model through a single network.  With the hierarchical approach, different types of features or feature representations can be utilized by each task.   %Besides, the direction for the class "something else" is often ambiguous and rather hard to annotate, which would introduce noise to the system. % when this class would be the only one present in the audio.
Therefore, we investigate a hierarchical classifier to first detect the presence of "speech" and "something else" classes using magnitude spectral features only. In the second level of the classification, spatial features are extracted and segments already detected to contain speech are further assigned with directional labels of "speech front" and "speech back", where the labels match the sound source direction with respect to the mobile array. 
The joint system offers a solution to the problem of differing tasks and features. On the other hand, the accuracy of the direction estimation does not only depend on its own performance anymore, since any missed speech samples from the sound event classifier would diminish its performance too. Fig.~\ref{fig:framework} depicts a high level overview of the proposed system.

\ssubsection{Sound Classification (Stage 1)}
\label{sec:speech}
%model design

\input detection.tex

\ssubsection{Localization (Stage 2)}
\label{sec:localization}
\input sloc.tex
%#model design, parameter choices?

\input baseline.tex

\vspace{-1mm}
\section{Evaluation}
\vspace{-1mm}

\label{sec:results}

\input tabledetection.tex

\input tablelocres.tex

\input tablejoint.tex

%\ssubsection{Speech Detection Results}
\paragraph*{Speech Detection Results}
%\label{sec:speechres}
\input results_detection.tex
\paragraph*{Localization Results}

\input locred.tex

%
%\ssubsection{Detection and Localization Results}
%\label{sec:jointres}
\paragraph*{Detection and Localization Results}

\input results_joint.tex

\vspace{-1mm}
\section{Conclusions}
\label{sec:conclusions}
%This work investigated the joint sound detection and localization (SELD) problem using a mobile microphone array with weak and source type dependent (hierarchical) label annotations. A two-stage hierarchical sound event detection and localization approach was proposed. %The performance was evaluated on a collected and a hierarchically annotated multi-channel mobile microphone array dataset.

This work proposes a two-stage hierarchical sound event detection and localization approach using a mobile microphone array. The first stage of the hierarchical model recognizes the sound event type, and the second stage is invoked for the direction estimation only for the blocks detected to contain speech. This structure allows the utilization of different types of features and network structures for the two stages, and accommodates the use of different hierarchy levels in the annotation of different sound classes. % The use of different feature time-scales for the tasks is not directly feasible with the non-hierarchical flat approach. 

The proposed method obtained better results in terms of average label score for every metric in contrast to a flat baseline classifier. % while both approaches showed strong performance in the novel weakly and hierarchically labeled mobile microphone array SELD task on a real recorded audio database.
The use of a mobile phone form factor microphone array and diverse real data pave way for future applications of SELD on practical mobile devices. 

%The sound event detection utilizes magnitude spectrum features and their temporal sequence with a convolutional recurrent neural network, while the direction classification determines the detected speech direction between the front and back using spatial features (sound propagation time delay and sound attenuation) between microphone signals with a similar type of neural network. The results favor the use of the hierarchical classification approach over a flat baseline.

In the future, the amount of sound events with direction labels could be increased to study the need for a class specific direction estimation. Similarly, a varying direction resolution for different classes could be investigated. 

%Addressing more directions than two or more classes could be explored in the future. 

%The classification was done with one second resolution. 
%For the sound event detection, the inclusion of the temporal information has been previously useful. For the detection of the direction, the CNN approach with temporal resolution decimation was used to result in a classifier that notices the presence of front and back directions, and is invariant of the sequence of front and back directions.

%\clearpage

\small{
\bibliographystyle{IEEEtran}
\balance{
\bibliography{refs19,refsAP}
} }

%\end{sloppy}
\end{document}

%% file: abstract.tex
Joint sound event localization and detection (SELD) is an integral part of developing context awareness into communication interfaces of mobile robots, smartphones, and home assistants. For example, an automatic audio focus for video capture on a mobile phone requires robust detection of relevant acoustic events around the device and their direction.
Existing SELD approaches have been evaluated using material produced in controlled indoor environments, or the audio is simulated by mixing isolated sounds to different spatial locations.  
This paper studies SELD of speech in diverse everyday environments, where the audio corresponds to typical usage scenarios of handheld mobile devices. In order to allow weighting the relative importance of localization vs. detection, we will propose a two-stage hierarchical system, where the first stage is to detect the target events, and the second stage is to localize them.

The proposed method utilizes convolutional recurrent neural network (CRNN) and is evaluated on a database of manually annotated microphone array recordings from various acoustic conditions. The array is embedded in a contemporary mobile phone form factor. The obtained results show good speech detection and localization accuracy of the proposed method in contrast to a non-hierarchical flat classification model.

%% file: introduction.tex
Sound source localization (SSL) aims to determine either the direction or position of the source in a continuous or discrete-valued space, and automatic sound event detection (SED) aims to recognize the classes of the source sounds present, and estimate their temporal activities.  
%to produce a symbolic description of detected sounds in the analyzed content.
 The SSL and SED have been extensively researched mostly as separate problems. Deep learning methods have brought improvements in SSL performance \cite{Xiao:ICASSP:2015,Chakrabarty:WASPAA:2017,Yalta:JRM:2017,Adavanne:EUSIPCO:2018, Vera-Diaz:MDPI:2018,Salvati:2018,VESPERINI201883,Pertila:2017:ICASSP,pertila:ICASSP:2019,Wang2018} and in SED~\cite{piczak2015environmental,takahashi2016deep,salamon2017deep,cakir2017convolutional} over traditional approaches. Recent approaches solving simultaneously the SED and SSL problems, i.e., joint sound event detection and localization (SELD) problem, include using \acp{CNN}~\cite{hirvonen2015classification,He2018:IS},   \ac{CRNN}~\cite{adavanne2018sound,cao2019polyphonic, politis2020overview}, and the Least Absolute Shrinkage and Selection Operator (LASSO)~\cite{8928942}. 
 % The best structure of SELD network is an open research question and weight transfer from a CNN or CRNN SED network to a similarly structured SSL network has been proposed~\cite{}. The work in~\cite{8928942} relies on Least Absolute Shrinkage and Selection Operator (LASSO) approach for binaural SELD instead of deep learning. 

As the research in the field has progressed towards machine learning approaches, the data used to train a system has a crucial impact on its performance. Larger and more diverse datasets enable learning more complex models that will generalize better to new conditions.
%A system can be expected to work well with data that is similar to the one that has been used for training. A large mismatch between training and testing data typically leads to significant drop in performance.
Since recording of acoustic scenes with spatiotemporal annotations is an extremely difficult task, there are no large-scale datasets of real scenes, and existing smaller ones are limited to evaluation of algorithms and are unsuitable for training deep-learning methods (e.g. LOCATA challenge \cite{evers2020locata}). That is in contrast, e.g., to automatic speech recognition where annotation is not a problem and recorded datasets exist with diverse range of conditions (e.g., the ASpIRE \cite{harper2015automatic} and CHiME \cite{barker2018fifth} challenges). Hence, for deep-learning based SSL researchers have generally relied on either simulations for training and testing on a small recorded dataset \cite{perotin2019crnn, nguyen2020robust}, or on emulated scenes with real recorded room impulse responses (RIRs). The second option allows integrating real acoustics with source signals of interest with a few RIR datasets available \cite{hadad2014multichannel, jeub2009binaural, stewart2010database}. Additionally, the SELD datasets related to the DCASE challenge have been generated with a large scale RIR collection from 15 rooms and a spherical microphone array \cite{politis2020overview}. The only annotated dataset for SELD with real recordings we are aware of is the one in \cite{brousmiche2020secl} for office environments.

 In this study, a microphone array embedded inside a flat mobile phone body was used to collect a dataset from diverse everyday environments to obtain insights for realistic mobile phone applications of the SELD approaches. To our knowledge, a microphone array in a mobile phone form factor has not been dealt with in previous SELD research. The flat microphone array shape imposes challenges to the spatial resolution rendering the task more difficult in contrast to, e.g., spherical arrays. Therefore, we do localization using only two categories, \emph{front} and \emph{back}, with respect to a mobile phone screen. This is motivated by possible mobile phone multimedia applications. % such as using the proposed method as a trigger for automatic audio focus emphasizing relevant sounds from a certain direction.
  We focus the evaluation on localization and detection of speech % because  of its importance in most applications,
  and detection of other prominent sounds of interest without localizing them.

State of the art SELD systems typically train a single system for joint localization and detection. In order to allow controlling the relative importance of localization vs. detection, we  propose a hierarchical approach of two separate deep neural networks, optimized for their corresponding tasks. %The detection part assigns "speech" and "something else" labels to each one second block. The localization network assigns a direction label "front" and "back" to each frame detected as speech.

%T  into two application motivated categories: {\emph front}, and \emph{back} with respect to a mobile phone screen
 %
 % 
 %The motivation is to reduce the annotation work while retaining a moderate temporal resolution.  

 %
%To investigate the proposed approach, a database was recorded and annotated with a time resolution of one second, referred as a block. This relaxed time resolution further reduces the cost of manual annotation, since each block is assigned multiple labels without more detailed timing information, also referred as tagging or weak labeling~\cite[Ch.~6]{virtanen2018computational}.However, the uncertainly of sound occurrence and location is limited to one second. % Each block is annotated with labels speech front, speech back, and {\it{else}} for anything non-speech other than background, i.e. interesting sound events such as music, horns, etc.   
 %
 %

%This paper describes the database collection, presents the sound source dependent weak labeling approach, addresses the SELD problem by a novel hierarchical two-stage deep neural network approach, and then compares the results to a baseline classifier.

In this paper Section~\ref{sec:data} describes the database collection. % setup and presents the sound source dependent weak labeling approach. 
Section~\ref{sec:method} describes the proposed hierarchical two-stage SELD approach. Section~\ref{sec:results} describes the evaluation of the SED, SSL, and the joint system and then compares the results to a baseline classifier. Finally,  Section~\ref{sec:conclusions} draws the conclusions and future directions.

%% file: dataset.tex
\input tabledata.tex

%\subsection{Data collection and annotation}
\paragraph*{Data collection and annotation}

%\hl{Aapo, your role is to briefly describe the recording setup (device setup (cameras, mics), data rate, annotation accuracy) for the real data only.

%You can use figures, but I’d limit them to two figures side-by-side. Maybe a picture of the recording device, and then an example picture of a video where providing a feel for the data. Data statistics, how many seconds of data we have in different environments.}

 Acoustic data for training and evaluation of the methods was collected in complex real-life environments. Actions and objects in the recordings aim to represent typical contents of casual videos recorded by a typical mobile phone user. The data set contains speech and other scene-specific sounds from common scenarios such as sports, moving vehicles and live music. In total there are 24 environment types including both indoor- and outdoor scenes, refer to Table~\ref{tab:environments}. %The environment types used in the recordings and the amounts of data (in 1-second blocks) for each environment are listed in Table~\ref{tab:environments}.
 The duration of the recordings varies from 10 to 180 seconds, and the total duration is 89~min. % duration of recorded audio data is 89~minutes.

\begin{figure}
     \centering
     \begin{subfigure}[b]{0.4\columnwidth}
     \includegraphics[width=\columnwidth,trim = {0mm 0mm 0mm 0mm}, clip]{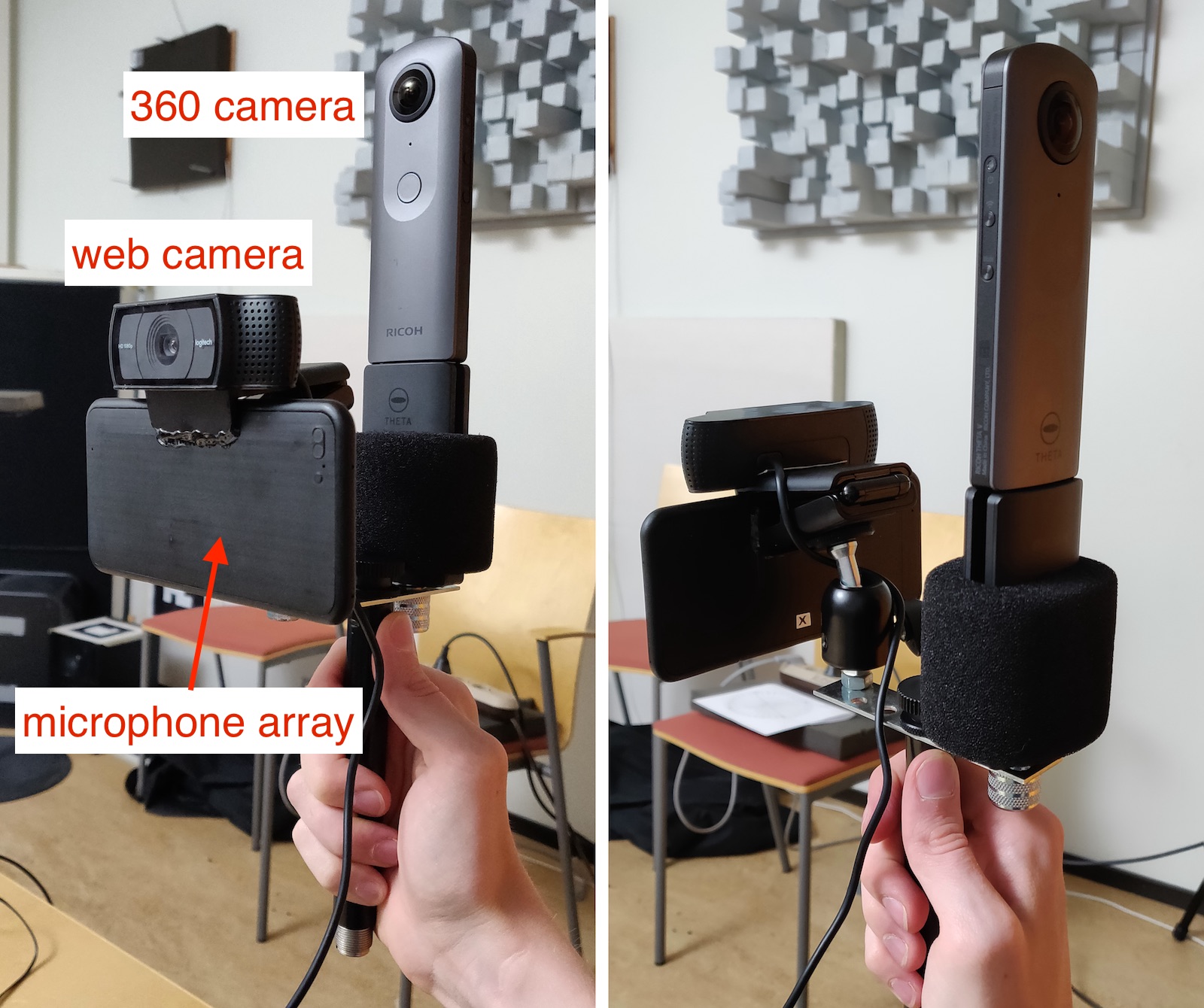}\vspace{-2mm}
     \caption{Pictures of the microphone array and other accompanying devices.\label{fig:arraypic}}
     
     \end{subfigure}
     \hfill
     \begin{subfigure}[b]{0.55\columnwidth}
      \centering
    \includegraphics[width=\columnwidth,trim = {50mm 22mm 35mm 33mm}, clip]{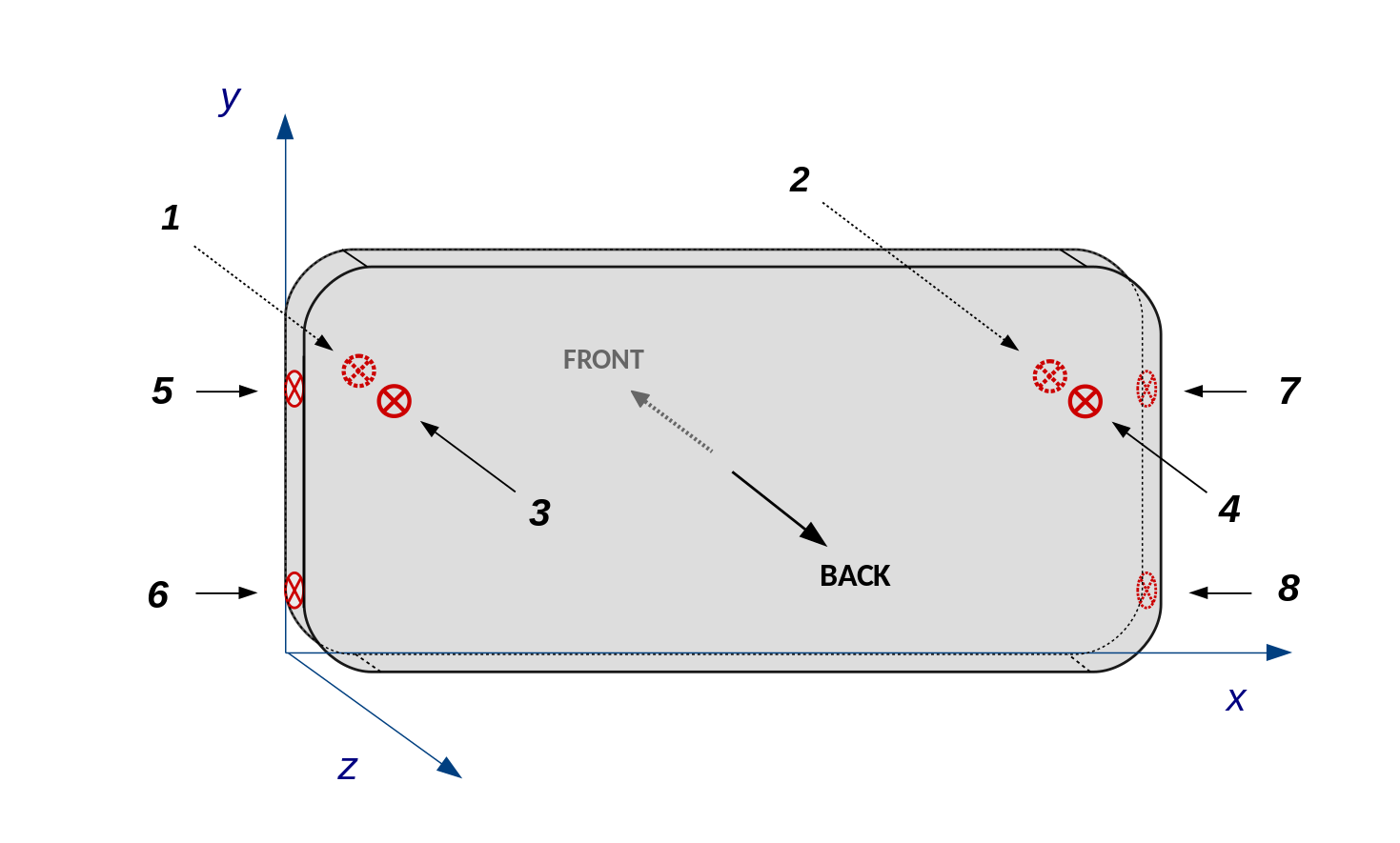}\vspace{-2mm}
    \caption{An illustration of the eight microphone placements and the directions "Front" and "Back" used to annotate the speaker direction. The x,y,z dimensions of the devices are: 140, 65, and 7~mm.\label{fig:array}}
    \end{subfigure}
\caption{Recording setup~(a), and microphone layout~(b).\label{fig:arraypictures}}
\vspace{-6mm}
\end{figure}

% Three recording devices were used in the data collection.
The audio data was collected with an eight-channel microphone array mounted to a custom 3D-printed rigid phone-shaped body. In addition, a 360$^\circ{}$ camera was used to make the annotation task easier and a web camera was used to collect video material for later research purposes. The devices were fixed to a hand-held microphone stand during the recordings that was was held from a grip below to prevent obstructing sensors. %the effect of microphones being covered by fingers.
 Figure~\ref{fig:arraypictures} illustrates the recording setup and the microphone array. 
 %
 %Refer to Fig.~\ref{fig:arraypic} for pictures of the recording setup, and Fig.~\ref{fig:array} for the microphone array.
 A laptop was used to collect the eight-channel audio as wav files at 48~kHz sampling rate and 32-bit resolution. 
%  The recording of the microphone array was controlled with a laptop and the audio files produced are in eight-channel wav-format with a sampling rate of 48 kHz and resolution of 32-bits.  %The placement of the microphones is illustrated in Fig.~\ref{fig:array}.

%\begin{figure}[t]
%    \centering
%    \includegraphics[width=.4\columnwidth,trim = {0mm 0mm 0mm 0mm}, clip]{arraypic.jpg}\vspace{-2mm}
%    \caption{Pictures of the microphone array and other accompanying devices.}
%    \label{fig:arraypic}
%    \vspace{-5mm}
%\end{figure}

The recorded signals were annotated with the following labels using one-second resolution: % The classes used in the annotations are %(1) A person holding the microphone array is speaking, (2) Other people near the array are speaking 
%  \begin{enumerate}[wide, labelwidth=!, labelindent=0pt,nosep, label={(\arabic*)}]
 (1)~{\bf Speech back}: A person was speaking from behind the microphone array, e.g., the person holding the device. 
(2)~{\bf Speech front}: A person in  front of the array was speaking.  
(3)~{\bf Something else}: An object was emitting interesting non-speech sound.
%\end{enumerate}
Multiple labels were allowed to be used simultaneously. The direction for the class "something else" was not used, since it was not considered as interesting as for the speech. Besides, the direction for this class is often ambiguous and rather hard to annotate.
%If no labels were assigned, the block was considered as background noise.

A sound was considered interesting if the sound source was  the focus of the recording or the sound was otherwise special in the context of the environment. For example, a scenario on a beach where a diver jumped into water in front of the array was assigned the label "Something else" during the time of a splash. Other swimmers talking in the background and creating similar splashing sounds were considered as background, since they were not in a key role from the perspective of the cameraman. Therefore, the boundary between interesting sounds and background noise is inevitably ambiguous, since the same sound can be assigned a label or considered as background noise depending on the situation. A general guideline used in the annotations was that when a sound source was close to the array and the sound was audible inside a block of data, it was assigned with the corresponding label. Other examples of  sounds labeled "Something else" include a guitar, a racing car, and a table tennis ball. All speech related sounds such as singing and whispering were assigned with labels "Speech back" and/or "Speech front". Multiple labels were assigned when multiple sound sources were present inside the one-second clip, including multiple sound sources of the same type. Audio blocks where a speech source is in the left or right directions % $\pm 30^\circ{}$
were omitted   due to ambiguous front/back direction. %Multiple classes can be active at the same time, including multiple sound sources of the same class.

%Directions of the annotated speech sounds were also labeled for each block of data. The directions used are front and back, determined by the location of the sound source with respect to the microphone array. 

%\begin{figure}[t]
%    \centering
%    \includegraphics[width=.6\columnwidth,trim = {50mm 22mm 35mm 33mm}, clip]{frontback.png}\vspace{-2mm}
%    \caption{An illustration of the eight microphone placements and the directions "Front" and "Back" used to annotate the speaker direction. The x,y,z dimensions of the devices are: 140, 65, and 7~mm.}
%    \label{fig:array}\vspace{-2mm}
%\end{figure}

%\ssubsection{
\paragraph*{Database for machine learning}
The audio was divided into six different folds. For the first five folds, the data was distributed so that each fold had the same proportion of \mbox{1-s} blocks captured in an inside and outside environment as the whole data set. Each recording was used as a whole in a single fold, and no recording was split between several folds. The sixth fold %was added later to include more
includes most data from speech sources simultaneously in front and back. 
Table~\ref{tab:classdir} displays the amount of \mbox{1-s}~blocks per each label class. It also includes counts for the cases where speech is present either back or front, and also back \emph{and} front.

\addtolength{\tabcolsep}{-4pt} 

\begin{table}[b]\vspace{-3mm}
    \centering
        \caption{Label appearance statistics for the folds. Note that any combination of labels can be present during a single block.}\vspace{-2.3mm}
    \label{tab:classdir}
    \scalebox{.87}{
    \begin{tabular}{c|c||c|c|c| |c|c|c }
    \toprule
   Fold & Total & Something & Speech Front  &  No & %\multicolumn{3}{c}{Direction classes for speech labels}
   Speech & Speech & Speech Front \\
     \#             & blocks & Else & or Back  &  labels &  Front only &  Back only &   and Back  \\
    \toprule

    1               & 871   & 376 & 376 & 213   &  187     & 177 & 12     \\
     
    2             & 748    &  327 & 310 & 218    &  61 & 243 & 6      \\
    
    3               & 768   & 187 & 383 & 305   &  98     &  271 & 14     \\
    
    4              & 892  & 292 &  407  & 327    &  173     & 204 & 30      \\
     
    5              & 719  & 376 &  201 & 307    &  24     &  176   & 1  \\
     
    6              & 1025  & 0 & 967 & 58    &  291     & 576 & 100      \\
    \end{tabular}
    } 

\end{table}
\addtolength{\tabcolsep}{4pt}

%% file: tabledata.tex
\ifoldr

\addtolength{\tabcolsep}{-4.5pt} 
\begin{table}[!b]
\vspace{-4mm}
    \centering \caption{Amount of recorded seconds for each environment type.}
    \label{tab:environments}\vspace{-2.3mm}
    \scalebox{0.65}{
    \begin{tabular}{ |c| c| c| c| c| c| c| c| c| c| c| c| c| c| c| c| c| c| c| c| c| c| c| c|  }
         \hline
        \rotatebox{90}{Apartment} &
        \rotatebox{90}{Industrial area} &
        \rotatebox{90}{Club room} &
        \rotatebox{90}{Studio room} 
        &\rotatebox{90}{Street} 
        &\rotatebox{90}{Meeting room} 
        &\rotatebox{90}{Corridor/office } &
        \rotatebox{90}{Park} &
        \rotatebox{90}{Live club} &
        \rotatebox{90}{Car} &
        \rotatebox{90}{Urban area} &
        \rotatebox{90}{Stairs} &
        \rotatebox{90}{Lake} &
        \rotatebox{90}{Cycle path} &
        \rotatebox{90}{Ship's deck} &
        \rotatebox{90}{Subway} &
        \rotatebox{90}{Store room} &
        \rotatebox{90}{Grocery store} &
        \rotatebox{90}{Marketplace} &
        \rotatebox{90}{Beach} &
        \rotatebox{90}{Harbour} & 
        \rotatebox{90}{Terrace} &
        \rotatebox{90}{Cafe} & 
        \rotatebox{90}{Terminal} \\
        \hline
           \rotatebox{90}{1094 }&
           \rotatebox{90}{920}&
           \rotatebox{90}{630}& 
           \rotatebox{90}{419}&
           \rotatebox{90}{404}&
           \rotatebox{90}{335}&
           \rotatebox{90}{245}&
           \rotatebox{90}{226}& 
           \rotatebox{90}{210}&
           \rotatebox{90}{185}&
           \rotatebox{90}{110}&
           \rotatebox{90}{105}& 
           \rotatebox{90}{65}&
           \rotatebox{90}{60}&
           \rotatebox{90}{60}& 
           \rotatebox{90}{51}& 
           \rotatebox{90}{45}&
           \rotatebox{90}{41}& 
           \rotatebox{90}{40}&
           \rotatebox{90}{20}&
           \rotatebox{90}{20}&
           \rotatebox{90}{15}& 
           \rotatebox{90}{15}& 
           \rotatebox{90}{11}\\
           \hline
    \end{tabular}
    } %\vspace{-4mm}
\end{table}
\addtolength{\tabcolsep}{4.5pt}

\fi

\addtolength{\tabcolsep}{-1pt} 
\begin{table}[!b]
\vspace{-4mm}
    \centering
        \caption{Duration (s) of recordings for each environment.}
    \label{tab:environments}\vspace{-2.3mm}
    \scalebox{0.8}{
    \begin{tabular}{ |l c| l c| l c|l c |    }
         \hline
        {Apartment} & 1094 &
       {Industrial area} & 920 &
        {Club room} & 630 &
        {Studio room}& 419  \\ \hline
        {Street} & 404 & 
        {Meeting room} & 335 &
        {Corridor/office} & 245 &
        {Park} &226 \\ \hline
        {Live club} &210 & 
        {Car} &185 &
        {Urban area} &110 &
        {Stairs} &105 \\ \hline
        {Lake} &65  & 
        {Cycle path} &60 &
        {Ship's deck} &60 &
        {Subway} &51 \\ \hline
        {Store room} &45 & 
        {Grocery store} &41 & 
        {Marketplace} &40 &
        {Beach} &20 \\ \hline
        {Harbour} & 20 & 
        {Terrace} &15 &
        {Cafe} & 15 &
        {Terminal} & 11 \\
           \hline
    \end{tabular}
    } %\vspace{-4mm}
\end{table}
\addtolength{\tabcolsep}{1pt}

%% file: detection.tex
In the sound classification stage, given a \mbox{1-s} block of audio, the aim is to recognize if there is speech and/or any other interesting sound events (\textit{i.e.} "something else" class) present at any time. The output of this stage is a probability value for both of these labels. Sound classification is done by feeding time-frequency acoustic features extracted from audio to a deep neural network that estimates the class probabilities. Feature extraction and classification are elaborated below.

%In the first step, time-frequency domain acoustic features are extracted from the audio. In the second stage, a \ac{DNN} maps the features into class probabilities. %based model is trained to do classification using the acoustic features as input, and the manually obtained sound annotations as target output. 

\begin{figure}[t]
    \centering
    \includegraphics[width=0.67\linewidth]{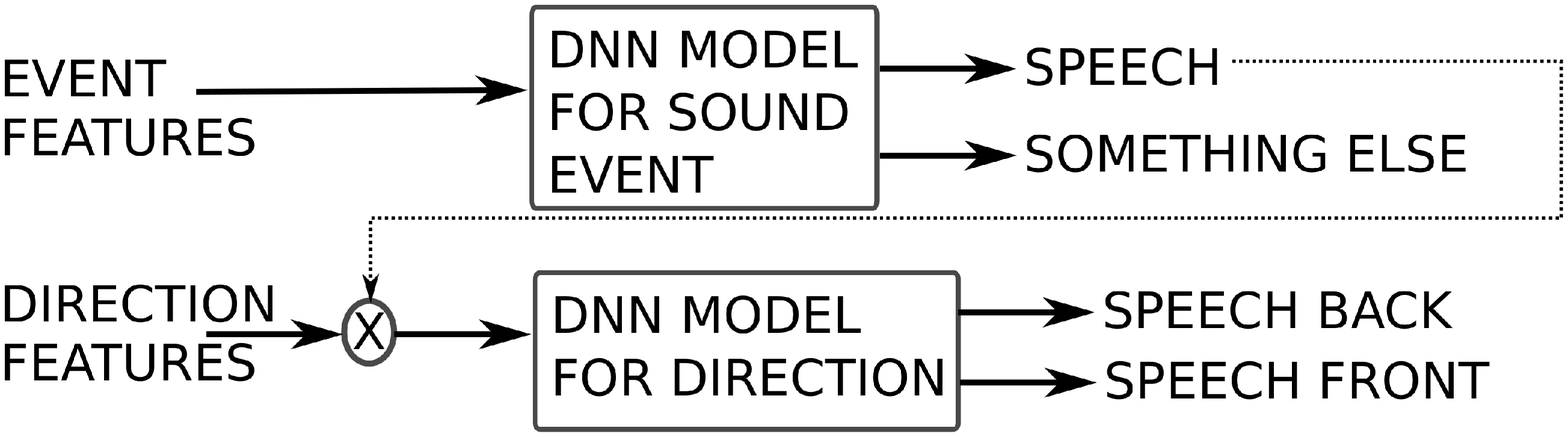}\vspace{-1mm}
    \caption{Joint classifier system framework for detection and localization of sound events. First stage detects the type of sound, second stage  estimates direction of detected speech.}
    \label{fig:framework}\vspace{-5mm}
\end{figure}

\ssection{Acoustic Feature Extraction}
\label{subsec:clsf_feat_ext}
%In the acoustic feature extraction stage, the given multi-channel audio is first combined, then log mel spectrogram is calculated and finally, the features are scaled in order to be fed to the DNN as input. 
%The feature extraction pipeline is presented in Figure~\ref{fig:feature_extraction}.

% \begin{figure}
%     \centering
%     \includegraphics[width=0.7\linewidth]{figs/classification/feature_extraction.eps}
%     \caption{Acoustic feature extraction pipeline for the sound event classification. The output shapes at several steps of the pipeline are given in brackets.}
%     \label{fig:feature_extraction}
% \end{figure}

%\paragraph*{Pre-processing step}
In a pre-processing step, the multi-channel audio is converted to mono by taking the average over the channels at each sample. The resulting monophonic audio signal is amplitude normalized by dividing with the maximum absolute value. % in the given one-second audio block. %The assumption for using monophonic audio is that the direction information, which can possibly be extracted from the channel differences, is  irrelevant to the sound classification. % In other words, we are rather interested in the content and not the direction of the sound. Nevertheless, experiments using multi-channel audio features have been conducted an reported in Section~\ref{sec:speechres}.

\paragraph*{Log-Mel Spectrogram Estimation}
The log-mel spectrogram features were obtained using 20 ms frames and 10 ms overlap in 40 mel frequency bands. The features are standardized to zero mean and unity variance using statistics from the training set.

\ssection{Deep Neural Network Model}\label{subsec:speechtrain}

The DNN technique utilized in the sound classification is a \ac{CRNN}. % CRNNs have theoretical advantages for sound classification in real-life environments, where sound events may exhibit wide variations in their frequency content and temporal structure.
Convolutional layers use small, shifting 2D kernels to extract higher level features that are invariant to local spectral and temporal variations. Recurrent layers are effective in modeling the longer term temporal context for the sound events. Combining convolutional and recurrent layers has been found suitable for various audio classification tasks such as sound event classification~\cite{cakir2017convolutional}, automatic speech recognition~\cite{amodei2016deep} and music genre classification~\cite{choi2016convolutional}.

The details of the CRNN architecture are as follows. Each 5-by-5 convolutional layer is followed by a rectified linear unit (ReLU) activation, and max-pooling by two in both time and frequency dimensions. The 3D output of the final convolutional layer is converted into 2D by reshaping the frequency and channel dimensions into a single dimension. This output is fed to one or multiple recurrent layers (long-short term memory~(LSTM)~\cite{hochreiter1997long} layers specifically). The LSTM output is fed to a fully connected feed-forward layer with logistic sigmoid activation that applies the same weights over each time step of the input. The resulting output at each time step are the two label probabilities of the i) speech presence using the merged "Speech Front" and "Speech Back" labels, and ii) "Something Else". Finally, max-pooling over time is applied over the sequence, and the two label probabilities are obtained for the one-second block of audio. The binary label predictions are obtained by using a threshold value of $0.5$ for the output probabilities.  

%The network training procedure is as follows. 
The network is trained to minimize the cross-entropy between the estimate output and the target output. Adam~\cite{Kingma:ICLR:2015} is used as the optimizer. After each epoch, the F1-score for the merged "speech" labels for the validation set is calculated. % If it is better than the best one from the earlier epochs, the model at that epoch is saved. 
 If the model does not improve for 25 epochs, the training is terminated. The best model based on the validation set score is used for testing.  In total, 197 different hyper-parameter combinations were evaluated with different amounts and configurations of the convolutional layers and the recurrent layers. The best model had 611k learnable parameters.

%% file: sloc.tex
 %\hl{Should this paragraph go to localization?}
 %The received microphone signal is a noisy mixture of source signals, each convolved with a \ac{RIR} $H_{i,n}\tfp$ between the $i$th microphone position and the $n$th source position. In the \ac{STFT} domain, the signal model can be written as
%\begin{equation}\vspace{-1mm}
%  \textstyle  X_i\tfp = \sum_n H_{i,n}\tfp S_n\tfp +  N_i\tfp,
%\end{equation}
%where $N_i\tfp$ is a mixture of interference sources, and additive noise that is independent between microphones, and $\omega_k=\jmath 2\pi k/K$ is the angular frequency with $K$ discrete DFT bins. 
%\hl{Maybe mention we combine front-left, front and front-right as "front" (and same for "back") due to practical reasons (low amount of data)}

The task of the localization step is to assign the labels of "Speech front" and "Speech back" to each 1-s block of audio, detected to contain speech in the first stage. %Refer to Fig.~\ref{fig:array} for an illustration of the array geometry and sound direction labels.
%
%The time sequence of directional labels of "speech front" and "speech back" is not assumed to be significant during the one second audio block, and only the appearance of either or both directions within the block is of interest. Therefore, 
%
%
%The CRNN architecture is  also used to solve the direction classification task by mapping a sequence of feature vectors into output probabilities or direction labels. % and processes them to output a two dimensional vector $\mathbf{o}=[o_1,o_2]^\textnormal{T}$ with values $o_i \in [0,1]$.
%
%The model architecture selected here is a , and the input is a set of concatenated feature values corresponding to one second input signal. %For a description of a \ac{CNN}, refer to e.g. \cite{chollet}.  The features with default parameter values include the pairwise \ac{TDOA} from channels (1,3) and (2,4), and they are of dimension $F \times P$, where $P=2$ is the number of pairs. The $F \times B$ magnitude features are obtained by averaging the magnitude differences of the same pairs. In total, there are $F\times (B+P)$ feature values per one second. 
%
 % The structure of the \ac{CNN} is listed in Table~\ref{table:cnn1}, and it %
 %
% 
%The magnitude spectrum features for the speech and "something else" classifier do not include spatial information about the speech source. However, s
%
% Spatial information about the source is required to classify its direction.
The speaker direction is assumed to be more stable compared to changes in the magnitude spectrum, and therefore longer 85~ms frames with 50~\% overlap are utilized. Two types of spatial features are extracted.

\ssection{\ac{TDOA} Feature}
The \ac{TDOA}, i.e. the sound propagation delay between a microphone pair, brings information about the dominating sound direction during each short processing frame. The case with speakers in front and back is evident by alternating \ac{TDOA} values. % between the speakers during the one-second block. 

%Since the task is to locate speech in front and/or back of the array, the microphone pairs that are maximally separated in this axis produce the most spatial information. 
 %
%They were estimated separately between two microphone pairs that had one of the microphones in the front and the other in the back surface of the device, and otherwise in the same (x,y) position, refer to Fig.~\ref{fig:array}.
%
%Therefore, two microphone pairs (1,3) and (2,4) located on the front and back surfaces on the left and right side of the device were used for TDOA estimation and are displayed in Fig.~\ref{fig:array}. 

 %
%The parameters (window length, amount of overlap, and window shape) of the feature extraction are defined in the \texttt{parameters} struct, and their values are illustrated in Table~\ref{table:defaultvalues}. The default window lenght is 4096 samples.
 %
% \hl{The default values are pairs (1,3) and (2,4), which correspond to forward and backward facing microphone pairs in the left and right side of the mobile array.%, refer to Fig~\ref{fig:arraygeometry}.
% 
 %
Since the microphone pair separation in the front-back axis is only 7~mm, the \ac{TDOA} resolution is limited to three possible values at 48~kHz. Therefore, \ac{TDOA} is obtained as maximum peak index of Fourier interpolated (factor of five) \ac{GCC}~\cite{Knapp:1976:TASLP} between microphones. 
\ssection{Magnitude Difference Feature}
Upon reaching the rigid device, the sound wave is partly reflected and partly diffracted, leading to frequency and angle dependent sound propagation effects. This is observed as a direction specific level difference between the microphones.
 %
% The attenuation of a sound wave reaching the microphones depends on the inclination angle and the dimensions of the mobile device with respect to each wavelength due to acoustic shadowing. 
% 
 The magnitude difference between the microphone pairs is used as the second spatial feature to capture this information: % It is defined and its values are illustrated in this subsection. 
%
%\begin{equation}
 $   D_{i,j}(b) =  \ln (M_i(b)) - \ln (M_j(b))$,
%\end{equation}
where $M_i(b)$ is the % mel-frequency resolution magnitude spectrum and $b$ refers to the $b$th mel-frequency band. The total number of used mel-bands is $40$. The feature value was similarly averaged over the pairs. 
  magnitude of the $b$th mel band of microphone $i$. The total number of  mel-bands was $40$. %The feature values are also % similarly obtained from and
%  averaged over the two microphone pairs (1,3) and (2,4). 

Features are obtained from two microphone pairs (1,3) and (2,4) located on the front and back surfaces on the left and right side of the device. The pairs are maximally separated in the dimension of interest, refer to Fig.~\ref{fig:array}. Both features are then averaged over the microphone pairs for robustness and to reduce the feature dimension by half.

 \paragraph*{\ac{DNN} Model for Localization}
As for sound event detection, the used model was a CRNN, where now two output values are related to  probabilities of speech front, and speech back. A similar training process and validation method was used as in Section~\ref{subsec:speechtrain}. %, except that three folds were used for training, two for validation, and one for testing. This is due to lack of directional diversity in a single validation set. Also, 
The Adam and Adamax~\cite{Kingma:ICLR:2015} %and adagrad~\cite{adagrad}
 optimizers were experimented with.
 %
%function grows near 0. Also, in the page \pageref{fig:frontback} 
 %
%\subsubsubsection{Training with Real Data}
%The section~\ref{sec:realdatacollection} describes the real-data collection method. This section describes, how the real-data was used to train the front/back classifier.
%
%
The training was done using k-fold-cross-validation with all one-second blocks containing speech from the folds listed in Table~\ref{tab:classdir}. Three of the folds were used for training, two for validation, and one for testing to guarantee sufficient number of directional labels in the validation set. % The table represents the multilabel information of the two classes using a categorical split into three separate classes, where the class "both" represents the number of blocks where "speech front" and "speech back" are simultaneously active, and "speech front" and "speech back" indicate the number of blocks  where only either label is present. 
%
 %The validation set was used to stop the training process by monitoring the decrease of the validation set's average F1-score over both direction labels. 
  The training was stopped if validation set's average F1-score over both direction labels started to decrease. 
Note that the case where both "Speech front" and "Speech back" are active is underrepresented, with only 6~\% of the samples belonging to this class. As a consequence, the speech samples are labeled as either with "Speech back" or "Speech front" labels, and the cases where both labels are active is effectively ignored. To address this data imbalance, a random oversampling strategy~\cite{VanHulse:2007} was applied to balance training and validation sets during training so that each unique combination of label values (i.e. class) would have the same amount of (partly repeated) training sequences. This reduced slightly the final F1-score in contrast to not using oversampling, but raised the performance of detecting speech in both directions. %Without considering the class imbalance, the network tries to propose a trivial solution, where
 %While this results in a good overall F1-score, it does not actually recognize simultaneous speech in front and back directions. 
 %
The best model had 170k parameters. 

%% file: baseline.tex
\ssubsection{Baseline}\label{sec:baseline}
The baseline comparison method is a flat CRNN classifier with three binary labels for each of the classes. The spatial features described in Section~\ref{sec:localization} are   only used, since i) the magnitude spectrum features could not be concatenated to spatial features due to the different frame lengths, and ii) the used spatial features already contain magnitude spectrum difference information. A similar training process as in Section~\ref{subsec:speechtrain} is used, but instead of speech-only F1-score, the weighted average F1-score over all three labels is used as the early stopping criteria. 
 The best model had 360k parameters.
%To prevent this, the error resulting from the class samples is weighted with the inverse value of their relative amount. As a result the samples in the "Both" class contribute an equal amount to the objective functions as the other two classes. The weighting value was empirically raised to the power 1.5 to further emphasize the effect. 

%% file: tabledetection.tex
\ifoldr
\begin{table}[b] 
\vspace{-5mm}
    \caption{Overall performance of the proposed CRNN method for sound classification.}
    \label{table:clsf_overall} 
    \centering\scalebox{1}{
    \begin{tabular}{c|c|c|c}
    & Precision & Recall & F1 (\%) \\
    \toprule
        Speech (Back or Front) &  89.3& 87.4& 88.4\\
        Something else & 76.6& 63.9& 69.7 \\
    \end{tabular}} 
\end{table}
\fi

\ifnewr\ifoldr
\begin{table}[b] 
\vspace{-5mm}
    \caption{Performance of the hirearchical CRNN method for sound classification.}
    \label{table:clsf_overall} 
    \centering\scalebox{1}{
    \begin{tabular}{c|c|c|c}
    & Precision & Recall & F1 (\%) \\
    \toprule
% Speech precision, recall, F1 score at thr 0.5: 0.8777, 0.8983, 0.8879
% Something else precision, recall, F1 score at thr 0.5: 0.7504, 0.6134, 0.6750

        Speech (Back or Front) &  87.8& 89.8& 88.8\\
        Something else & 75.0 & 61.3& 67.5 \\
    \end{tabular}} 

\end{table}

% Speech front precision, recall, F1 score at thr 0.5: 0.8702, 0.8877, 0.8788
% Speech back precision, recall, F1 score at thr 0.5: 0.7204, 0.7801, 0.7491
% Something else precision, recall, F1 score at thr 0.5: 0.7110, 0.6075, 0.6552

\fi\fi

\ifoldr

\addtolength{\tabcolsep}{-2pt} 
\begin{table}[t]
\caption{Confusion matrix of the proposed \ac{CRNN} method for sound detection. The instances where both classes are present are treated as a separate class for the visualization.}
    \label{fig:clsf_overall} 
\newcommand\items{4}   %Number of classes
\arrayrulecolor{white} %Table line colors
\noindent
\scalebox{0.9125}{
\begin{tabular}{cc*{\items}{|c}|}
\multicolumn{1}{c}{} &\multicolumn{1}{c}{} &\multicolumn{\items}{c}{Predicted label} \\ \hhline{~*\items{|-}|}
\multicolumn{1}{c}{} & 
\multicolumn{1}{c}{} & 
\multicolumn{1}{c}{\rot{Nothing}} & 
\multicolumn{1}{c}{\rot{Speech}} & 
\multicolumn{1}{c}{\rot{Else}} &
\multicolumn{1}{c}{\rot{Both}} \\ \hhline{~*\items{|-}|}
\multirow{\items}{*}{\rotatebox{90}{Actual label}} 
&Nothing  & 1114   & 140  & 164 & 10   \\ \hhline{~*\items{|-}|}
&Speech  & 203   & 1727  & 18 & 108   \\ \hhline{~*\items{|-}|}
&Else  & 302   & 76   & 523 & 50\\ \hhline{~*\items{|-}|}
&Both  & 35   & 143   & 76 & 334 \\ \hhline{~*\items{|-}|}
\end{tabular}%
}\hspace{-4mm}%
\renewcommand\items{4}   %Number of classes
\arrayrulecolor{white} %Table line colors
\noindent
\scalebox{0.9125}{
\begin{tabular}{cc*{\items}{|E}|}
\multicolumn{1}{c}{} &\multicolumn{1}{c}{} &\multicolumn{\items}{c}{Predicted label} \\ \hhline{~*\items{|-}|}
\multicolumn{1}{c}{} & 
\multicolumn{1}{c}{} & 
\multicolumn{1}{c}{\rot{Nothing}} & 
\multicolumn{1}{c}{\rot{Speech}} & 
\multicolumn{1}{c}{\rot{Else}} &
\multicolumn{1}{c}{\rot{Both}} \\ \hhline{~*\items{|-}|}
\multirow{\items}{*}{\rotatebox{90}{ }} 
&Nothing & 78.0& 9.8& 11.5& 0.7 \\ \hhline{~*\items{|-}|} 
&Speech  &  9.9& 84.0& 0.9& 5.3 \\ \hhline{~*\items{|-}|} 
&Else  &31.8& 8.0& 55.0& 5.3 \\ \hhline{~*\items{|-}|} 
&Both  & 6.0& 24.3& 12.9& 56.8 \\ \hhline{~*\items{|-}|} 
\end{tabular}
}
\addtolength{\tabcolsep}{2pt} 
\vspace{-5mm}
\end{table}

\fi

\ifnewr

\addtolength{\tabcolsep}{-3pt} 
\begin{table}[t]\vspace{-0mm}
\caption{The confusion matrix of SED (proposed).}
%\caption{The confusion matrix of the proposed hierarchical \ac{CRNN} method for speech and else class detection. } % The instances where both classes are present are treated as a separate class for the visualization.}
    \label{fig:clsf_overall} \vspace{-2mm}
\newcommand\items{4}   %Number of classes
\arrayrulecolor{white} %Table line colors
\noindent
\scalebox{0.895}{
\begin{tabular}{cc*{\items}{|c}|}
\multicolumn{1}{c}{} &\multicolumn{1}{c}{} &\multicolumn{\items}{c}{Predicted label} \\ \hhline{~*\items{|-}|}
\multicolumn{1}{c}{} & 
\multicolumn{1}{c}{[frames]} & 
\multicolumn{1}{c}{{Nothing}} & 
\multicolumn{1}{c}{{Speech}} & 
\multicolumn{1}{c}{{Else}} &
\multicolumn{1}{c}{{Both}} \\ \hhline{~*\items{|-}|}
\multirow{\items}{*}{\rotatebox{90}{True label}} 
% model 143 
% Normalized confusion matrix
% [[0.76890756 0.1092437  0.11344538 0.00840336]
% [0.08706226 0.84484436 0.00680934 0.06128405]
% [0.33964248 0.11041009 0.48895899 0.06098843]
% [0.04761905 0.23639456 0.08163265 0.63435374]]
% 2020-01-27 02:30:25.818117
&Nothing  & \bf1098   & 156  & 162 & 12   \\ \hhline{~*\items{|-}|}
&Speech  &179 & \bf1737 &  14 & 126   \\ \hhline{~*\items{|-}|}
&Else  & 323 & 105 & 465  & 58 \\ \hhline{~*\items{|-}|}
&Both  &28 & 139 & 48 & \bf373\\ \hhline{~*\items{|-}|}
\end{tabular}%
}\hspace{-4mm}%
\renewcommand\items{4}   %Number of classes
\arrayrulecolor{white} %Table line colors 
\noindent
\scalebox{0.9}{
\begin{tabular}{cc*{\items}{|E}|}
\multicolumn{1}{c}{} &\multicolumn{1}{c}{} &\multicolumn{\items}{c}{Predicted label} \\ \hhline{~*\items{|-}|}
\multicolumn{1}{c}{} & 
\multicolumn{1}{l}{[\%]} & 
\multicolumn{1}{c}{\hspace{-5.5mm}{Nothing}} & 
\multicolumn{1}{c}{\hspace{-3.3 mm}{~Speech}} & 
\multicolumn{1}{c}{\hspace{-2 mm}{Else}} &
\multicolumn{1}{c}{{Both}} \\ \hhline{~*\items{|-}|}
\multirow{\items}{*}{\rotatebox{90}{ }} 
&Nothing & 76.9  & 10.9 &  11.3  &  0.8 \\ \hhline{~*\items{|-}|} 
&Speech  &  8.7  & 84.5 &   0.7  &  6.1  \\ \hhline{~*\items{|-}|} 
 &Else   &  34.0 & 11.0 &  48.9  &  6.1 \\ \hhline{~*\items{|-}|} 
&Both    &  4.8  & 23.6 &   8.2  & 63.4  \\ \hhline{~*\items{|-}|} 
\end{tabular}
}
\vspace{-3mm}
\end{table}
\addtolength{\tabcolsep}{3pt} 

\fi

%%%%%%%%%%%% BASELINE %%%%%%%%%%%%%%%%%
\ifnewr

\addtolength{\tabcolsep}{-3pt} 
\begin{table}[t]
\caption{The confusion matrix of SED (baseline).}
%\caption{The confusion matrix of the flat baseline \ac{CRNN} method for (merged) speech and else class detection.}% The instances where both classes are present are treated as a separate class for the visualization.}
    \label{fig:clsf_overall_baseline} \vspace{-2mm}
\newcommand\items{4}   %Number of classes
\arrayrulecolor{white} %Table line colors
\noindent
\scalebox{0.9}{
\centering
\begin{tabular}{cc*{\items}{|c}|}
\multicolumn{1}{c}{} &\multicolumn{1}{c}{} &\multicolumn{\items}{c}{Predicted label} \\ \hhline{~*\items{|-}|}
\multicolumn{1}{c}{} & 
\multicolumn{1}{c}{[frames]} & 
\multicolumn{1}{c}{{Nothing}} & 
\multicolumn{1}{c}{{Speech}} & 
\multicolumn{1}{c}{{Else}} &
\multicolumn{1}{c}{{Both}} \\ \hhline{~*\items{|-}|}
\multirow{\items}{*}{\rotatebox{90}{~True label}} 
&Nothing  & 1037 & 194 & 173 &  24   \\ \hhline{~*\items{|-}|}
&Speech  &266 &1575 &  38 & 177   \\ \hhline{~*\items{|-}|}
&Else  &210 & 152& \bf 511 &  78 \\ \hhline{~*\items{|-}|}
&Both  &40 & 166 & 104 & 278\\ \hhline{~*\items{|-}|}
\end{tabular}%
}\hspace{-4mm}%
\renewcommand\items{4}   %Number of classes
\arrayrulecolor{white} %Table line colors 
\noindent
\scalebox{0.89}{
\begin{tabular}{cc*{\items}{|E}|}
\multicolumn{1}{c}{} &\multicolumn{1}{c}{} &\multicolumn{\items}{c}{Predicted label} \\ \hhline{~*\items{|-}|}
\multicolumn{1}{c}{} & 
\multicolumn{1}{l}{[\%]} & 
\multicolumn{1}{c}{\hspace{-5.5mm}{Nothing}} & 
\multicolumn{1}{c}{\hspace{-3.3 mm}{~Speech}} & 
\multicolumn{1}{c}{\hspace{-2 mm}{Else}} &
\multicolumn{1}{c}{{Both}} \\ \hhline{~*\items{|-}|}
\multirow{\items}{*}{\rotatebox{90}{ }} 
&Nothing &  72.6 &  13.6 &  12.1 &   1.7 \\ \hhline{~*\items{|-}|} 
&Speech  &  12.9 & 76.6  &  1.8  &  8.6 \\ \hhline{~*\items{|-}|} 
 &Else   &  22.1 &  16.0 & 53.7 &   8.2 \\ \hhline{~*\items{|-}|} 
&Both    &  6.8 &  28.2 &  17.7 &  47.3 \\ \hhline{~*\items{|-}|} 
\end{tabular}
}
\vspace{-6mm}
\end{table}
\addtolength{\tabcolsep}{3pt} 

\fi

%% file: tablelocres.tex
\ifoldr
\addtolength{\tabcolsep}{-1pt} 
\begin{table}[t]
\caption{Confusion matrices of speech localization computed over the folds. Left table shows  accumulated results, right table shows relative results. The instances where both classes are present are treated as a separate class for the visualization.}\label{table:resdir}
\newcommand\items{3}   %Number of classes
\arrayrulecolor{white} %Table line colors
\noindent
\centering
\scalebox{1}{
\begin{tabular}{cc*{\items}{|c}|}
\multicolumn{1}{c}{} &\multicolumn{1}{c}{} &\multicolumn{\items}{c}{Predicted label} \\ \hhline{~*\items{|-}|}
\multicolumn{1}{c}{} & 
\multicolumn{1}{c}{} & 
\multicolumn{1}{c}{\rot{Back}} & 
\multicolumn{1}{c}{\rot{Front}} & 
\multicolumn{1}{c}{\rot{Both}} \\ \hhline{~*\items{|-}|}
\multirow[t]{\items}{*}{\rotatebox{90}{\hskip -25pt Actual label}} 
&Back  & 1508   & 18  & 121   \\ \hhline{~*\items{|-}|}
&Front  & 21   & 672  & 141   \\ \hhline{~*\items{|-}|}
&Both  & 22   & 52   & 89  \\ \hhline{~*\items{|-}|}
\end{tabular}%
}
%\hspace{2cm}%
\arrayrulecolor{white} %Table line colors
\noindent
\scalebox{1}{
\begin{tabular}{cc*{\items}{|E}|}
\multicolumn{1}{c}{} &\multicolumn{1}{c}{} &\multicolumn{\items}{c}{Predicted label} \\ \hhline{~*\items{|-}|}
\multicolumn{1}{c}{} & 
\multicolumn{1}{c}{} & 
\multicolumn{1}{c}{\rot{Back}} & 
\multicolumn{1}{c}{\rot{Front}} & 
\multicolumn{1}{c}{\rot{Both}} \\ \hhline{~*\items{|-}|}
&Back  & 91.6   & 1.1  & 7.3   \\ \hhline{~*\items{|-}|}
&Front  & 2.5   & 80.6  & 16.9   \\ \hhline{~*\items{|-}|}
&Both  & 13.4   & 31.9   & 54.6  \\ \hhline{~*\items{|-}|}
\end{tabular} 
}
\vspace{-5mm}
\end{table}
\addtolength{\tabcolsep}{1pt} 

\fi

\ifnewr 
\addtolength{\tabcolsep}{-2pt} 
\begin{table}[b]\vspace{-2mm}
\caption{Speech localization confusion matrix (proposed). \label{table:resdir}}%The instances where both classes are present are treated as a separate class for the visualization.}
\vspace{-3mm}
\newcommand\items{3}   %Number of classes
\arrayrulecolor{white} %Table line colors
\noindent
\centering
\scalebox{0.90}{
\begin{tabular}{cc*{\items}{|c}|}
\multicolumn{1}{c}{} &\multicolumn{1}{c}{} &\multicolumn{\items}{c}{Predicted label} \\ \hhline{~*\items{|-}|}
\multicolumn{1}{c}{} & 
\multicolumn{1}{c}{[frames]} & 
\multicolumn{1}{c}{{Front}} & 
\multicolumn{1}{c}{{Back}} & 
\multicolumn{1}{c}{{Both}} \\ \hhline{~*\items{|-}|}
\multirow[t]{\items}{*}{\rotatebox{90}{\hskip -20pt True label}} 
&Front  & 676 & 20 &138\\ \hhline{~*\items{|-}|}
&Back  & 23     & 1558    & 66    \\ \hhline{~*\items{|-}|}
&Both  &    58 & 28     & 77   \\ \hhline{~*\items{|-}|}
\end{tabular}%
\vspace{-2mm}
}
%\hspace{2cm}%
\arrayrulecolor{white} %Table line colors
\noindent
\scalebox{.85}{
\begin{tabular}{cc*{\items}{|E}|}
\multicolumn{1}{c}{} &\multicolumn{1}{c}{} &\multicolumn{\items}{c}{Predicted label} \\ \hhline{~*\items{|-}|}
\multicolumn{1}{c}{} & 
\multicolumn{1}{c}{[\%]} & 
\multicolumn{1}{c}{{Front}} & 
\multicolumn{1}{c}{{Back}} & 
\multicolumn{1}{c}{{Both}} \\ \hhline{~*\items{|-}|}
&Front &  81.1  & 2.4   & 16.5    \\ \hhline{~*\items{|-}|}
&Back  &  1.4   & 94.6  & 4.0    \\ \hhline{~*\items{|-}|}
&Both  &  35.6  & 17.2  & 47.2  \\ \hhline{~*\items{|-}|}
\end{tabular} 
}
\vspace{-5mm}
\end{table}
\addtolength{\tabcolsep}{2pt}

%% file: tablejoint.tex
\ifoldr
\arrayrulecolor{black} %Table line colors
\addtolength{\tabcolsep}{-2pt} 
\begin{table}[hb]
\vspace{-5mm}
    \centering
    \caption{Speaker direction detection performance (a): assuming the speech classifier works \%100 correct, (b): for the samples where sound classifier detects speech presence, (c): for the samples where speech classifier \textit{correctly} detects speech presence.}
    \label{table:joint}    
    \scalebox{1}{
\begin{tabular}{cc|c|c|c}
    &  & Precision & Recall & F1 \\
    \toprule
     (a) & Front   &  87.3& 95.7& 91.3\\
      & Back  & 91.5& 96.1& 93.7 \\
    \midrule
     (b)&  Front  &  83.9& 97.3& 90.1\\
      &  Back & 79.6& 95.8& 86.9 \\
\midrule
     (c) &  Front  &  89.6& 97.3& 93.3\\
       & Back   & 91.7& 95.8& 93.7 \\
    \end{tabular}
    } 
\end{table}
\addtolength{\tabcolsep}{2pt} 
\fi

\ifnewr
\arrayrulecolor{black} %Table line colors
\addtolength{\tabcolsep}{-3pt} 
\begin{table}[t]
\vspace{-0mm}
    \centering
    \caption{Speaker direction and else class detection performance (A): proposed hierarchical approach, (B): baseline  }\vspace{-2mm}
    \label{table:joint}    
    \scalebox{1}{
\vspace{-1mm}
\scalebox{0.9}{
\begin{tabular}{cc|c|c|c|c}
    & Label & Accuracy \%& Precision \%& Recall \%& F1 \%\\
    \toprule
%Speech front precision, recall, F1 score at thr 0.5: 0.8432, 0.8957, 0.8687
%Speech back precision, recall, F1 score at thr 0.5: 0.7932, 0.8436, 0.8177
%Else precision, recall, F1 score at thr 0.5: 0.7504, 0.6134, 0.6750
%Accuracy score for speech front 0.952219788971, speech back 0.852478598447, else 0.819032450727

     (A) & Speech front  & 95.2 &  84.3& \bf 89.6& 86.9\\
      & Speech back  & \bf 85.2 &\bf 79.3& \bf 84.4& \bf 82.8 \\
      & Something else & \bf 81.9 & \bf 75.0& 61.3& \bf 67.5 \\
    %\midrule
    & Average (unweighted) & \bf 87.4 & \bf 79.5 & \bf 78.4 & \bf 79.1 \\
    \midrule
%Speech front precision, recall, F1 score at thr 0.5: 0.9234, 0.8706, 0.8962
%Speech back precision, recall, F1 score at thr 0.5: 0.7583, 0.7486, 0.7534
%Something else precision, recall, F1 score at thr 0.5: 0.7021, 0.6309, 0.6646
% Accuracy score for speech front 0.959984073263, speech back 0.823412303404, else 0.80489747163

(B)&  Speech front  & \bf 96.0 & \bf 92.3 &  87.1 & \bf 89.6 \\
      &  Speech back & 82.3 &75.8 & 74.9 & 75.3 \\
      &  Something else &80.5 &70.2 & \bf 63.1 & 66.5 \\
    %\midrule
    & Average (unweighted) & 86.3 & 79.4 & 75.0 & 77.1 \\
    \end{tabular}}
    } \vspace{-6mm}
\end{table}
\addtolength{\tabcolsep}{3pt} 
\fi

%% file: results_detection.tex
% TABLE MOVED BEFORE 
% results for the speech and something else classification
% This section describes the classification results for the speech detection.

%\ssection{Overall performance}

%\begin{figure}
%    \centering
%    \includegraphics[width=0.6\linewidth]{figs/detection/test_311_fold_avg.png}
%    \caption{Confusion matrix of the proposed CRNN method. The instances where both %classes are present are treated as a separate class for the visualization.}
%    \label{fig:clsf_overall}
%\end{figure}

%The overall performance of the proposed \ac{CRNN} method for sound classification is presented in Table~\ref{table:clsf_overall}. 
The confusion matrix for the two-stage \ac{CRNN} speech classification model is given in Table~\ref{fig:clsf_overall} and the baseline results for comparison with speech direction outputs merged into a single class are given in  Table~\ref{fig:clsf_overall_baseline}. The instances where both classes are present are treated as a separate class for the visualization. The sample numbers inside each box are obtained by % binarizing the test set predictions with the threshold 0.5, and
 accumulating the binary output of the test fold values of the six folds using  k-fold cross-validation. %\textit{AE: why present confusion matrix first and only then overall results? Would first present overall results, then go deeper into confusion matrix and probably explain a bit about the confusions.}
 The percentage values represent the fraction of the predicted labels assigned for each audio block with the corresponding true label. 
 
 The hierarchical approach has better classification performance in almost all classes ("Silence", "Speech", "Both") except for the "Else"  class in contrast to the baseline. 
 
%The informal listening tests and output visualizations point to the fact that a significant amount of the errors for classifying speech happen around the onset and offset points. These are often ambiguous regions where it is quite hard even for humans to pinpoint the exact onset and offset points. In addition, the 

The "Something else" performance is quite low in both approaches compared to the class "Speech". This can be attributed to i) the labeling ambiguity problem, and ii)  to the scarcity of data, which makes it hard to capture the characteristics of all the various types of sound events that are included in this class.

%% file: locred.tex
Table~\ref{table:resdir} depicts the confusion matrix for the different location classes during frames with annotated speech. The results are obtained by accumulating the binary test folds label predictions over the six folds (i.e. k-fold cross-validation). The direction classifier is described in Section~\ref{sec:localization}. 
 %
%Based on a grid search, the best network parameters were
%a CNN kernel size of $3\times3$ with 32 feature maps, 
%128 neurons in the fully connected layers, 
%an $\ell_2$-regularization value of 0.01 in the CNN and fully connected layers, and 
%a dropout probability $p=0.25$ before the output layer. The adagrad optimizer and a mini-batch size of 64 samples were used.  
%
%
 The speech emitted from the back direction was more accurately recognized than speech from the front. This is expected, since all the samples of a speaker holding the device are labeled with "Speech back". Since they were emitted close to the array, they inherently had a better SNR in contrast to the front direction, which contained more distant talkers. The cases where both the directions were active are detected with the least accuracy. This can be attributed to having the least amount (only 6~\%) of data from such cases. 

%\textit{AE: also here, please consider presenting overall results first and then confusion matrices}
%Macro and micro averaged F1-scores are 0.676 and 0.911 respectively.

%peech front precision, recall, F1 score at thr 0.5: 0.9184, 0.9478, 0.9329
%peech back precision, recall, F1 score at thr 0.5: 0.9165, 0.9519, 0.9339
%inal F1 score = 0.933519202229. (Average over both labels)
%onfusion matrix, without normalization
%[   0    0    0    0]
%[   0  676   20  138]
%[   0   23 1558   66]
%[   0   58   28   77]]
%ormalized confusion matrix
%[       nan        nan        nan        nan]
%[0.         0.81055156 0.02398082 0.16546763]
%[0.         0.01396478 0.94596236 0.04007286]
%[0.         0.35582822 0.17177914 0.47239264]]

%%%%%%%%% baeline 

\fi

%% file: results_joint.tex
The performance for the joint sound classification and direction detection system is presented in Table~\ref{table:joint}. In the hierarchical approach, the speaker direction is estimated only when there is speech detected by the sound classifier. For frames without speech detection, the  "Speech front" and "Speech back" labels are set to zero.

The proposed hierarchical classifier has the highest performance in terms of the F1-score for the labels "Speech back" and "Something else". In contrast, the "Speech front" is overall better detected with the flat model. This is most likely attributed to the observed  difficulty in detecting the presumably more distant and thus weaker speech signals in front of the array. Since the hierarchical system only passes the frames labeled as speech, the performance is deteriorated by the errors accumulating from the two separate classification steps. The "Speech back" detection capability for the hierarchical model is significantly higher (7.5~\% points  higher in terms of F1-score) than that of the baseline, rendering the performance of the hierarchical model better in terms of overall performance. This is also evident in the unweighted average performance values, which are all higher for the proposed model in contrast to the baseline.